\documentclass[11pt]{article}
\pdfoutput=1
\usepackage{jcappub,natbib}
\input{colordvi.tex}

\def\ga{\mathrel{\raise.3ex\hbox{$>$\kern-.75em\lower1ex\hbox{$\sim$}}}}
\def\la{\mathrel{\raise.3ex\hbox{$<$\kern-.75em\lower1ex\hbox{$\sim$}}}}

\def\gsim{\mbox{\raisebox{-.6ex}{~$\stackrel{>}{\sim}$~}}}

\def\rt{{\tilde r}}
\def\be{\begin{equation}}
\def\ee{\end{equation}}
\def\beq{\begin{equation}}
\def\eeq{\end{equation}}
\def\beqa{\begin{eqnarray}} 
\def\eeqa{\end{eqnarray}}

\usepackage[utf8]{inputenc}

\usepackage{graphicx,amsmath,amsfonts,amssymb,slashed}
\usepackage{bbold,wasysym}

\usepackage{amsmath,amsfonts,bbm,euscript}

\usepackage{xcolor}
\usepackage{color}
\usepackage{hyperref}

\hypersetup{colorlinks, citecolor=blue, linkcolor=red, urlcolor=blue}

\title{\LARGE {Properties of Ultralight Bosons from Heavy Quasar Spins via Superradiance} 
}

\author[a, b]{Caner \"Unal,}
\author[c, d]{Fabio Pacucci,}
\author[c, d]{Abraham Loeb}

\affiliation[a]{CEICO, FZU -- Institute of Physics of the Czech Academy of Sciences,\\
Na Slovance 2, 182 21, Prague, Czech Republic}
\affiliation[b]{Department of Physics, Ben-Gurion University, Be’er Sheva 84105, Israel}
\affiliation[c]{Center for Astrophysics, Harvard \& Smithsonian,
Cambridge, MA 02138, USA}
\affiliation[d]{Black Hole Initiative, Harvard University,
Cambridge, MA 02138, USA}

\abstract{\\
The mass and the spin of accreting and jetted black holes, at the center of Active Galactic Nuclei (AGNs), can be probed by analyzing their electromagnetic spectra. For this purpose, we use the Spin-Modified Fundamental Plane of black hole activity, which non-linearly connects the following four variables (in the source frame): radio luminosity, X-ray or optical luminosity (via the [OIII] emission line), black hole mass and spin. Taking into account the uncertainties in luminosity measurements, conversion factors, relativistic beaming and physical properties of the AGN system, we derive lower bounds on the spins of a group of heavy, jetted AGNs. Using these results, we study the direct implications on the mass spectrum of the ultra-light particles of scalar (axion-like), vector (dark photon) and tensor types (additional spin-2 particles). We close unexplored gap in the parameter space $10^{-20}-10^{-19}$eV. We obtain upper bounds on the axion decay constant (equivalently lower bounds on the self-interaction strength) considering self-interactions could prevent the axion particles entering the instability, and be the reason for non-observation of superradiance. Assuming axion/scalar is described by mass and decay constant, we obtain upper limits on what fraction of dark matter can be formed by ultra-light particles and find that single spieces axion-like light particle can constitute at most $10\%$ of the dark matter in the mass range: $ 10^{-21} < \mu \, (\mathrm{eV}) <  10^{-17}$. Moreover, we derive similar bounds for vector and spin-2 particles and find that  light vector fields can constitute at most $10^{-6}$  of the dark matter in $10^{-21}\, \mathrm{eV} < \mu < 10^{-17} \, \mathrm{eV}$ range, and light spin-2 fields  can constitute at most $10^{-9}$   of the dark matter in $10^{-23}\, \mathrm{eV} < \mu < 10^{-17} \, \mathrm{eV}$ range.
}

\emailAdd{unalx005@umn.edu}
\emailAdd{fabio.pacucci@cfa.harvard.edu}
\emailAdd{aloeb@cfa.harvard.edu}

\begin{document}
 \maketitle
\flushbottom

\section{Introduction}

The spin is one of the three fundamental physical quantities that characterize a classical black hole (BH), the others being its mass and electric charge. The BH spin plays an important role in a number of astrophysical processes including accretion, as well as producing relativistic jets \cite{refgeodesic,refnovikovthorne,BZoriginal,Ruderman:1975ju,Wald:1974np,Penrose:1969pc,Penrose:1971uk,Pacucci:2020orw} (for a recent review see \cite{jetsAGNsreview}). Two well-established methods \cite{BHspinsummary}, namely X-ray reflection and continuum fit, use the spectrum of the emitted radiation to measure the spin of stellar-mass and super-massive BHs. In addition to these methods, the direct imaging of the volume immediately around the event horizon of nearby super-massive black holes, with the Event Horizon Telescope (EHT), can provide crucial information on the spin \cite{m87spin}. However, in order to accurately obtain the spin, both the reflection spectrum and the continuum fit methods require very high signal-to-noise data, while EHT-like techniques can only be applied to nearby Active Galactic Nuclei (AGNs). In fact there are only about $\sim 20$ accurate spin measurements thus far \cite{BHspinsummary}. The present work, together with \cite{refsmfp}, aims at determining spins of a large number of AGNs in a more accurate way.

There are significant implications from theory, simulations and data  suggesting that the jet power is linked with the BH spin \cite{BZoriginal,Moderski:1998ak,BZderive2,refkomissarov,Fanidakis:2009ct,Garofalo:2009ki,Narayan:2011eb,Tchekhovskoy:2012bg} (See also  \cite{jetspincorrelation,Garofalo:2014tga,daly16,Schulze2017}). Early theoretical results estimate that jet power/efficiency has a quadratic dependence on the spin. This quadratic dependence exists for slow rotations, but as spin value increases the functional dependence increases nonlinearly with higher powers of spin value according to various simulations \cite{largespindependence1,largespindependence2} and recently a step towards theoretical understanding of these results have been performed in \cite{refsmfp}. Furthermore, the predictions from the theoretical and simulations have also been confirmed by observational data \cite{refsmfp}. This fact implies that the significant differences in jet power/efficiency could result from the different spin values of the central BH. Here, we provide further evidence for this hypothesis. 

For this purpose, we first started by adding spin as a variable to the relation, called Fundamental Plane of BH activity (FP), describing accreting and outflowing BHs \cite{kordingfp2006,Plotkin:2011dy,fp1,fp2,fp3,fp4,flatspectrum}. FP describes jet and disk symbiosis \cite{jetdiskcoup} via X-ray luminosity, radio luminosity and BH mass. Therefore, we describe the active BH system with 4 variables in \cite{refsmfp}, $L_{radio}-L_{Xray}-M-a$ which is called as Spin Modified Fundamental Plane (SMFP)  \footnote{FP is describes accreting and jetting BH pheonemena with two independent quantities, $\{M,{\dot M}\}$, and we describe it with three independent physical quantities, $\{M, {\dot M},a\}$.}. However there exist uncertainties related with the origin of the X-ray radiation. Moreover, different relativistic effects for different frequencies further decrease the accuracy of formulations. Therefore, \cite{Saikia:2015ega} suggested and used the [OIII] emission line instead of X-ray luminosity in FP and called it Optical Fundamental Plane. We take Optical FP as our starting point and include spin as an additional variable since we already know it has a crucial role in both accretion outflow mechanisms. We find that in our new formulation, all data points stay within remarkable proximity to this new 4-variable plane, $L_{\rm radio}-L_{\rm OIII}-M-a$, where quantities indicate radio luminosity, OIII line luminosity, mass of BH and its spin value, respectively.

In this work, we also improve our results by considering relativistic corrections which can considerably alter the interpretation of physical parameters. We employ the bulk Lorentz factor of the jet ($\Gamma_j$) and the angle of deviation of the jet from the line of sight ($\theta_j$) as we test and employ our Spin Modified Fundamental Plane (SMFP) relation. We used NGC 4151, 3C 120, M87, MRK 79 whose $\Gamma_j$ and $\theta_j$ values are known accurately so that our predictions match with data even better with respect to \cite{refsmfp} which neglected these corrections. Also in the predictive part we employ AGNs whose $\Gamma_j$ and $\theta_j$ are relatively well-measured. This allows us to compute luminosities in the source frame and improved the accuracy of our estimations. For the uncertainties, we estimate various types of errors and we incorporated them to find conservative lower bounds on the jetted AGN spins.

We explore the implications of our results on the particle physics. The Ultra-Light Bosons (ULB) are possibly contributing to the mass budget of our universe. They have attracted significant attention as they might answer some of the puzzles in cosmology and they are well-motivated in particle-physics point of view \cite{Arvanitaki:2009fg,Arvanitaki:2010sy,Hu:2000ke,Hui:2016ltb} (see a review \cite{Marsh:2015xka}). There has been a considerable amount of work investigating the implications of their existence on cosmological and astrophysical observables. Some of these signatures are on the fluctuations of the potential that modify the arrival time of pulsar signals \cite{refptaulb}, Lyman-$\alpha$ forest \cite{lymanaconstraints}, superradiance of black holes \cite{SRreview}, galactic rotation curves \cite{fuzzyrotation}, star clusters \cite{effectsoffluctuationsongalaxy}, SMBHs\cite{Annulli:2020lyc}  and binary pulsars (for scalars \cite{Blas:2016ddr} and for spin-2 \cite{Armaleo:2019gil}) (see also, e.g., \cite{gravulbreview}).

We focus on the production of ULB degrees of freedom via superradiance, a phenomenon that  exists as the Compton wavelength of the light particle of mass, $\mu$, is comparable to the horizon size of the black hole, $G \times M$, where $G$ is the Newton's constant of gravitation. If a BH rotates, then the particles can extract the rotational energy of the BH from the ergosphere region. This instability can produce large amounts of such particles relatively fast with respect to astrophysical processes and deplete the spin of the BH. In our analysis, we close the gap that was unexplored \cite{Davoudiasl:2019nlo}, $10^{-20}-10^{-19}$eV and by combining with previously measured AGN spins, our AGN spin calculations, we probe the largest parameter space to date. We found that if there is no or negligible self/external interaction, then considerable fraction of the phenomenologically interesting parameter space is ruled out. We obtain the following results:
\begin{itemize}
\item $2.9 \times 10^{-21} \, {\rm eV} \, - \,  1.7 \times 10^{-17} \, {\rm eV}$ for scalars (spin-0 particles),
\item $6.4 \times 10^{-22} \, {\rm eV} \, - \,  1.7 \times 10^{-17} \, {\rm eV}$ for vectors (spin-1 particles),
\item $2.6 \times 10^{-23} \, {\rm eV} \, - \,  1.7 \times 10^{-17} \, {\rm eV}$ for tensors (spin-2 particles).
\end{itemize}

These constraints are valid if the self/external interactions of the light degrees of freedom are too weak. For example, if self-interactions are strong enough, this removes the light particle out of the instability regime and prevents the production of superradiance \cite{nosuperradiancebyinteractions1,nosuperradiancebyinteractions2}. Therefore, one can derive constraints on the strength of self/external interactions by using the non-observation of superradiance. This strength is typically indicated by the decay constant, the axion scale, $f_a$. The decay constant is inversely proportional to the interaction strength, hence the results can be interpreted in two ways: i) if superradiance is not observed due to self-interactions, then we can derive upper bounds on the decay constant, implying larger self-interactions, and ii) if superradiance is observed then this implies a small self-interaction (large decay constant), hence we can derive lower bounds for the decay constant.

This paper is organized as follows: In Section \ref{sec:spinsmfp} we discuss the correlation between accretion and jet power for BHs, their spin dependence, called SMFP, and how to extract the spin value from multi-wavelength observations. By estimating several sources of errors, we obtain the conservative lower bounds for the spin values. Section \ref{sec:reviewsuperradiance} is devoted to a brief summary of superradiance and Section \ref{sec:constraints} is probing the properties of ultra-light particles of scalar, vector and tensor (spin-2) types. We constrain the mass spectrum for each type and further obtain upper limits on the axion decay constant, $f_a$, for scalar (axion-like) particles via the strength of the self-interactions that can restrain the superradiance phenomena. In the final part, we investigate what fraction of dark matter can be in the form of ultra-light particles in the mass range $10^{-21}-10^{-17}$ eV. We conclude in Section \ref{sec:conc} and discuss the implications of our results. We set $\hbar=1, c=1$, and $M_p \equiv 1/\sqrt{8 \, \pi \, G}\simeq 2.43\times 10^{18} \,$ GeV.

\section{Determining the BH Spin Parameter Using Multi-Wavelength Data}
\label{sec:spinsmfp}

Radio power from BHs is typically associated with collimated jets \cite{jetradiorelation}, while the origin of the X-ray emission is linked to the accretion disk, to the corona \cite{xraysfromcorona} and to the jet as well. Since jet power is also related to the accretion rate, radio and X-ray luminosities are expected to be correlated \cite{flatspectrum,kordingfp2006,Plotkin:2011dy,fp1,fp2,fp3,fp4} (see also \cite{fundamentalplane1,fundamentalplane2,fundamentalplane3}).  However, several studies \cite{flatspectrum,fp1,fp2} have suggested that there are multiple sources for X-rays. The uncertainty regarding the dominant contributor to the X-ray emission leads to a challenge in both estimating the source-frame X-ray luminosity (the Lorentz factor of the region releasing X-rays) and its corresponding mechanism. In the hard/low state of X-ray binaries, the emission is thought to be dominated by the jet \cite{Plotkin:2011dy}, whereas in the presence of large accretion flows X-rays are expected to be produced mainly by the corona \cite{Kaaret:2017tcn}. Moreover, the corresponding Lorentz factors can, in principle, be highly frequency dependent.

 In order to bypass these uncertainties, the forbidden [OIII] line luminosity, which is non-relativistic and nearly isotropic and informative about both the accretion rate and mass of the black hole, is chosen as a variable instead of X-ray power in FP equation \cite{Saikia:2015ega}. As the rest-frame [OIII] line is in the optical range ($\lambda = 500.7$ nm), the correlation of radio-[OIII]-M has been referred to as the Optical Fundamental Plane of BH activity.

\begin{figure}
\centering{ 
\includegraphics[width=0.75\textwidth]{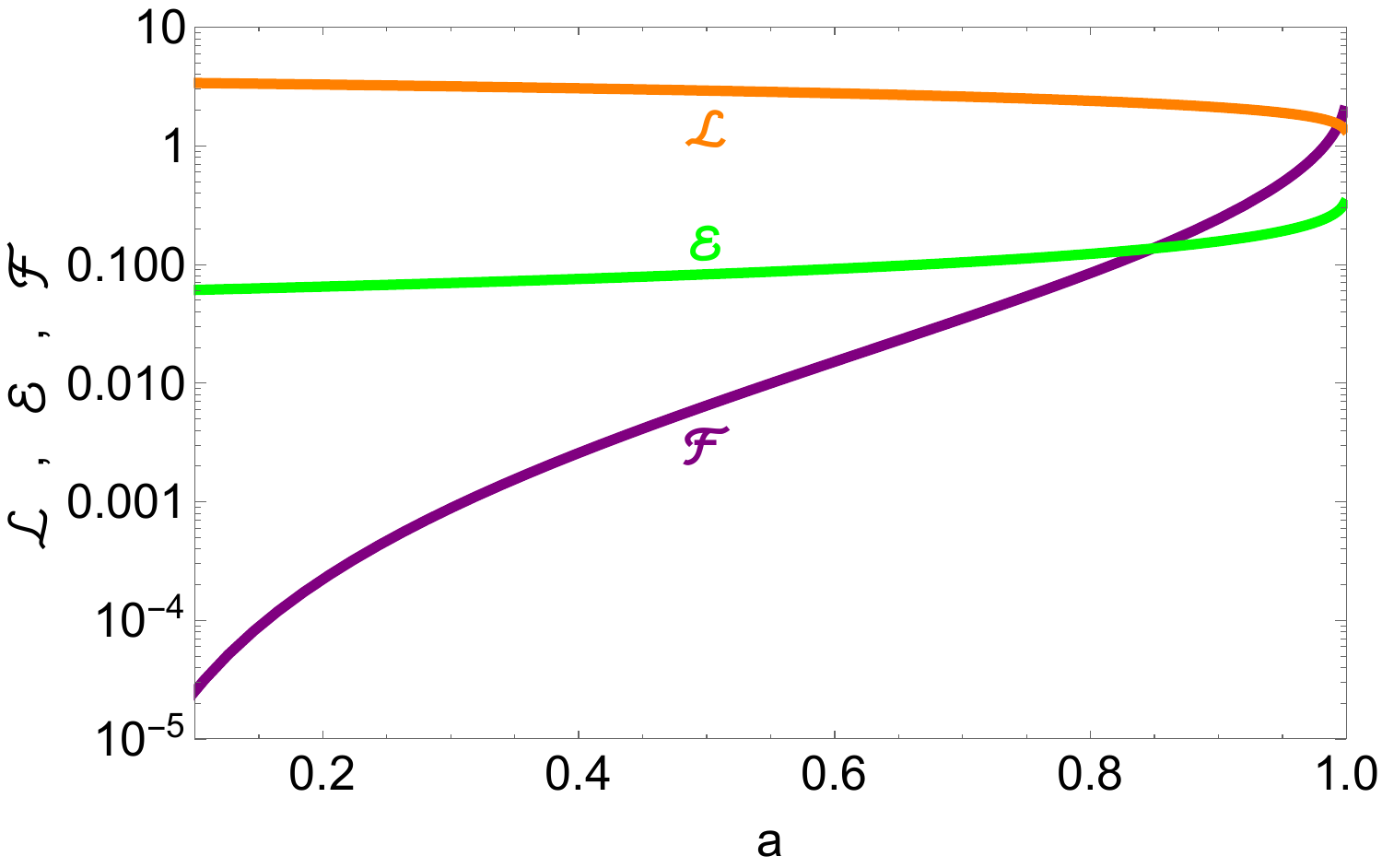}}
\caption{Spin dependencies of i)$\cal{F}$, spin dependent function of fundamental plane given in \eqref{eqcalF},  ii)${\cal L}$, the specific angular momentum given in \eqref{eqradeff}, and  iii)${\cal E}$, radiation efficiency given in \eqref{eqspinmoddisk}.}
\label{figspindependence}
\end{figure}

Theoretical studies, simulations and data agree that for moderately and rapidly rotating BHs, the dominant jet contribution is powered by the extracted rotational energy, via the Blandford-Znajek process \cite{BZoriginal,Moderski:1998ak,BZderive2,refkomissarov,luminosityaccretionrate,Sikora:2006xz,Fanidakis:2009ct,Garofalo:2009ki,Narayan:2011eb,Tchekhovskoy:2012bg} (See also  \cite{jetspincorrelation,Garofalo:2014tga,daly16,Schulze2017,Rusinek:2020djn,Soares:2020haa}). Without spin and magnetic field coupling \cite{mhdinstability1,mhdinstability2}, one can produce jets with lower power \cite{BPoriginal}). On the other hand, the accretion disk \cite{Accretiondiskbasics,alphadiskmodel, AGN_vs_star, Lens_2019} is the main driver of the bolometric luminosity. Jet and bolometric luminosities are poised to have a different dependence on the spin. As shown in Fig. \ref{figspindependence} and in \cite{refsmfp}, the jet power grows faster than the accretion power growth with spin. This will result in a deviation in the ratio of jet and accretion power, which allows us to extract the spin information from a given AGN. SMFP relation predicts that any data approximately lies on the 4-dimensional radio-OIII-M-$a$ plane, hence  by providing the information about radio, OIII and M, one can immediately find the AGN spin parameter when the propagation and beaming effects are taken into account.

In \cite{refsmfp}, the statistical analysis including 10 AGNs has shown that the data lies closer to the the Spin Modified plane with respect to the original Fundamental Plane equation. The standard deviation in a typical sample using the standard FP is about 1 dex, while in \cite{refsmfp} it is found to be about 0.5 dex. It is worth noting that the spin values of the 10 AGNs were determined independently and relatively accurately by conventional measurements, and they are used as an inherent physical charge of the physical charge of the BH system, ``not"  a free parameter to improve the analysis.

\begin{center}
\begin{table}[t]
\hspace{-0.8cm}
\begin{tabular}{|c|c|c|c|c|c|c|}  \hline       
Object & \;\;  $M/10^6 M_\odot$  \;\; & \;\; $ \log L^{obs}_{radio}  $  \;\; &  \;\;  $ \log L^{obs}_{OIII} $  \;\; & \;\;  $a$ \;\; & \;\;  $\Gamma_j$ \;\;   & \;\;  $\Theta_j$ \\ \hline 
 NGC  4151 &  $ 45 $ & $40.3$  & $41.48$ & $0.94^{+0.058}_{-0.04}$ \cite{Keck:2015iqa}   & 1.85 \cite{Beckmann:2005fg} & $40^{\circ}$ \cite{Winge:1999fy}   \\ \hline
3C 120 &  $55$ & $42.48$  & 41 &  $0.966^{+0.032}_{-0.016}$  \cite{Lohfink:2013uwa} & 5.5 \cite{Hovatta:2008mj} & $9.7^{\circ}$ \cite{Hovatta:2008mj}   \\ \hline
 MRK  79 &  52  & 39  & 41.3   & $0.7\pm 0.1$\cite{Gallo:2010mm} & $\sim 1$ \cite{Gallo:2010mm} & $24^{\circ}$ \cite{Gallo:2010mm}   \\ \hline
 M87 &  $6500$ & $41.18$  & 38.9 & $0.9\pm 0.1$ \cite{m87spin2} & 4.6 \cite{Walker:2018muw} & $17^{\circ}$ \cite{Walker:2018muw}   \\ \hline
\end{tabular}
\caption{The AGNs that are used to calibrate the SMFP. Luminosities are in units of $erg \, s^{-1}$. Spin, $\Gamma_j$ and $\Theta_j$ have been extracted from the references indicated. Mass values are obtained from Ref \cite{largespincatalog} except M87 (for that we use \cite{m87mass}), while radio and [OIII] luminosities are obtained from the NED archive whenever available.
}
\label{tabcalibration}
\end{table}
\end{center}

Now, we incorporate the spin dependence as described in \cite{refsmfp}. Additionally, we also include relativistic effects on the observational data by employing bulk Lorentz boost factor ($\Gamma_j$) and viewing angle ($\theta_j$), which allow us to write SMFP equation with radio-OIII-M-$a$ variables as
\begin{equation}
{\rm log}_{10} \, L^{obs}_{radio} / \delta_j^2 - 0.83 \, {\rm log}_{10} \, L^{obs}_{OIII} -0.82 \, {\rm log}_{10}\, M  = {\rm log}_{10}{\cal F}(a) + {\cal D} \,
\label{eqsmfp}
\end{equation}
where $L^{source}_{radio}=L^{obs}_{radio}/ \delta_j^2$ and $\delta_j$ is given in \eqref{eqforboost}, and ${\cal D}\simeq 0.1$ is set by the minimization of the $\chi^2$ error according to the data given in Table \ref{tabcalibration} and $a$ is the dimensionless spin parameter defined as $J/GM^2$. $\cal{F}$ shows the spin dependence of the FP relation and it is given by 
\begin{equation} 
{\cal F}(a) =  \left [ \, a^2 \, {\cal L}^2 \, {\cal E}^4 \, \right]^{1.42} \, /  \, \, {\cal E}^{\,0.83} \, ,
\label{eqcalF}
\end{equation}
where ${\cal F}(a=0) \to 0$ and ${\cal F}(a=1) \to 3.07$.

Radiation efficiency, ${\cal E}$,  can be expressed as in \citep{refgeodesic}
\begin{equation}
{\cal E} (a) = 1 - \frac{\rt^{3/2} - 2 \rt^{1/2} \pm a }{\rt^{3/4} \left( \rt^{3/2} - 3 \rt^{1/2} \pm 2 a \right)^{1/2}} \; \;  \bigg|_{\rt = {\tilde r}_{ISCO}} \;,
\label{eqspinmoddisk}
\end{equation}
where $\rt = r  / GM$. This equation gives ${\cal E} (a=0) \simeq 0.057$ and ${\cal E} (a=1) \simeq 0.423$.

 ${\cal L} $ is the scaled specific angular momentum and it is obtained by using the geodesic expression at the equator as \citep{refgeodesic}
\begin{equation}
{\cal L} \equiv   \frac{\left( \rt^2 \,  \mp \,  2a \rt^{1/2} \, +\, a^2 \right)}{\rt^{3/4} \, \left( \rt^{3/2}  \, - \, 3 \rt^{1/2} \, \pm \, 2a  \right)^{1/2}} \;  \bigg|_{\rt = {\tilde r}_{ISCO}} ,
\label{eqradeff}
\end{equation}
where upper/lower signs are for prograde/retrograde orbits. Note that at large large distances, $(r/GM)\gg1$, we have ${\cal L} \simeq \rt^{1/2}$ which is the standard Keplerian result.  ${\cal L} (a=0) =2\sqrt{3} $ and ${\cal L} (a\to1) =1.15$. We visualized $\cal{F}, {\cal L}, {\rm and} \,  {\cal E}$  in Fig. \ref{figspindependence}.

 For 4 AGNs given in Table \ref{tabcalibration} we have a detailed information, with the help of them the spin modified fundamental plane relation can describe data better and we obtain standard deviation as low as 0.3. This result has two sources: i)inclusion of relativistic corrections ii) formulation in OIII instead of X-ray. However, we should also note that we have fewer AGNs this time in our analysis since having detailed information about AGN systems(radio, OIII, M, boost factor, viewing angle) is not very common. We use Table \ref{tabcalibration} AGNS to calibrate the SMFP. We also note that one can have a more precise statistical analysis by also allowing to vary the coefficients of X-ray luminosity and BH mass, which requires a larger sample, but such a sample can not be used for precision predictions. Therefore, we employ the coefficient predictions from a larger analysis \cite{Saikia:2015ega} (see also \cite{Saikia:2016blk}). In order to get more precision, we focus on AGNs which are studied in detail so that on top of their mass and spin, their Lorentz boost and viewing angle is determined relatively precisely.

 In the next step, we will employ SMFP in special quasars such that we can predict their spin value relatively precisely, and derive reliable lower bounds. These heavy quasars are called Flat Spectrum Radio Quasars (FSRQ), which have a nearly scale invariant power spectrum at radio frequencies. They are blazars hence they send their jets in a small angle deviation from line of sight which improves our predictions considerably since in such a case relativistic effects are only determined by the Lorentz boost factor. Finally, since we see FSRQ typically directly, the conversion between OIII line and bolometric luminosity keep their linear relation until highest luminosity values.

To compute radio and [OIII] line luminosities in the source frame, we implement the following procedure. \\
\textbf{Radio Luminosity:} We select 5 GHz as the reference frequency for our calculation. For each AGN we use in our model, we search for radio-frequency measurements in the literature. Whenever a radio flux is not available at 5 GHz, we assume a flat spectrum such that $L_{\mathrm{radio}} \propto \nu$, so that we calculate the corresponding radio flux, and then luminosity, at 5 GHz.\\
\textbf{[OIII] Line Luminosity:} We measure the forbidden [OIII] line luminosity ($\lambda_{\mathrm{[OIII]}} = 500.7$ nm) using bolometric corrections to transform from the bolometric luminosity. Although a direct measurement of the flux from a spectrum would provide a better estimate for the [OIII] line luminosity, in most cases the spectral energy distribution around $\lambda_{\mathrm{[OIII]}}$ is not publicly available. In fact, the highest redshift for which the [OIII] line falls within the optical/NIR spectrum is $z=0.85$, while most of the sources we investigate here are at $z > 1$. We use the BH mass and the Eddington ratio (defined here as $L_{\mathrm{bol}}/L_{\mathrm{Edd}}$, where $L_{\mathrm{bol}}$ is the bolometric luminosity of the source and $L_{\mathrm{Edd}}$ is its Eddington luminosity) provided in \cite{ghis1,ghis2}, to calculate the bolometric luminosity of the source, assumed to be isotropically emitted. In order to map the bolometric luminosity to the [OIII] luminosity, we use the bolometric corrections presented in \cite{Pennell:2017gov}. In particular, we use the linear bolometric correction:
\begin{equation}
    \log_{10}(L_{\mathrm{OIII}}) = \log_{10}(L_{\mathrm{bol}}) -  (3.532 \pm 0.059)
\end{equation}
as it provides a better fit to data with $\log_{10}(L_{\mathrm{bol}}) > 46$, which is the case for almost all the AGNs in our sample.

\begin{center}
\begin{table}[t]
\centering
\begin{tabular}{|c|c|c|c|}  \hline       
Object & \;\;  $M/10^6 M_\odot$  \;\;  & \;\;  $a$  \;\;   & \;\; Ref \;\; \\ \hline 
 $^1$  1H0707-495 & $3\pm1$  & $>0.97$ &   \cite{Done:2015yqa}  \\ \hline
$^2$ IRAS13224–3809 & $6.3$  & $>0.987$ &   \cite{largespincatalog}  \\ \hline
$^3$ MRK 335 & $14.2 \pm 3.7$  & $>0.91$ &   \cite{largespincatalog} \\ \hline
$^4$ Mrk 110 & $25.1 \pm 6.1$  &  $ >0.89$ &   \cite{largespincatalog}  \\  \hline
$^5$ NGC 4151  & $45.7^{+5.7}_{-4.7}$ & $>0.9$  &  \cite{largespincatalog} \\ \hline
$^6$ Ark 120 & $118 \pm 16$  & $>0.85$ &   \cite{recentspinref} \\ \hline
$^7$ 1H0419-577 & $\sim 340$ &$>0.96$  &   \cite{h0419ref} \\ \hline 
M87 & $\sim 6.5 \times 10^3$ & $0.9\pm 0.1$  &  \cite{m87spin2}   \\ \hline  
\end{tabular}
\label{tabmeasured}
\caption{Source name, mass and spins for the BHs considered in this work, along with the corresponding reference for the data used. The BHs in this table corresponds to "measured" category whose spin is determined by conventional methods.}
\end{table}
\end{center}

\begin{table}[t]
\hspace{-1cm}
\begin{tabular}{|c|c|c|c|c|c|c|}  \hline       
Object & \;\;  $M/10^6 M_\odot$  \;\; & \;\; $L^{obs}_{radio} $ \, [GHz]  \;\; &  \;\;  $L^{obs}_{OIII} $  \;\;  & \;\; $\Gamma_j$ \;\; & \;\; $\theta_j$ \;\; & \;\;  $a$  \;\; \\ \hline

$^8$    B2 0218+35    & $4\times10^2$   &   \;\; $43.1$ (0.074)  \;\; &  \;\; $41.2$   \;\; &  16   &  $3^{\circ}$  & $\geq 0.99$  \\ \hline
$^{(*)}$  $^{8}$  B2 2155+312    & $4\times10^2$   &   \;\; $43.2$ (0.074)  \;\; &  \;\; $41.5$   \;\; &  12.2   &  $3^{\circ}$  & $\geq 0.99$  \\ \hline
$^8$    PMN 2345+155    & $4\times10^2$   &   \;\; $44.4$ (20)  \;\; &  \;\; $42.0$   \;\; &  13   &  $3^{\circ}$  & $\geq 0.73$  \\ \hline
$^{9}$ PKS 0142-278    &  $6\times10^2$   & $43.4$ (0.080) &  $42.0$  & 12.9 &  $3^{\circ}$   & $\geq 0.97$  \\ \hline
$^{(*)}$ $^{9}$ PKS 1055+018   &  $6\times10^2$   & $43.5$ (0.080) &  $42.4$  & 12. &  $3^{\circ}$   & $\geq 0.94$  \\ \hline
$^{9}$ TXS 0716-332   &  $6\times10^2$   & $43.2$ (0.074) &  $42.1$  & 12.2 &  $3^{\circ}$   & $\geq 0.94$  \\ \hline
$^{9}$ PKS 1057-79    &  $6\times10^2$   & $44.9$ (20) &  $42.2$  & 11 &  $3^{\circ}$   & $\geq 0.79$  \\ \hline
$^{10}$ 4C 55.17 0954+556       & $1\times 10^3$   &   $43.2$ (0.038) & $41.3$ &   13  &  $2.5^{\circ}$  & $\geq 0.99$     \\ \hline
$^{(*)}$ $^{10}$ PKS 0528+134 & $1\times 10^3$   &   $46.6$ (15) & $43.3$ &   13  &  $3^{\circ}$  & $\geq 0.92$     \\ \hline
$^{10}$  PKS 0048-071     & $1\times 10^3$   &   $43.6$ (0.074) & $42.8$ &   15.3  &  $3^{\circ}$  & $\geq 0.87$     \\ \hline
$^{10}$ DA55 0133+47       & $1\times 10^3$   &   $45.4$ (15) & $42.6$ &   13  &  $3^{\circ}$  & $ \geq 0.80$     \\ \hline
$^{11}$  GB6 J0805+6144 & $1.5 \times 10^3$  &  $45.6$ (5) & $42.9$ &  14   &  $3^{\circ}$    & $ \geq 0.86$       \\ \hline
$^{(*)}$ $^{11}$  S4 0820+560 & $1.5 \times 10^3$  &  $43.7$ (0.151) & $43.0$ &  14   &  $3^{\circ}$    & $\geq 0.74$       \\ \hline
$^{11}$ PHL 5225 2227-088 & $1.5 \times 10^3$  &  $43.0$ (0.080) & $42.9$ &  12   &  $3^{\circ}$    & $ \geq 0.69$       \\ \hline
$^{12}$ PKS 0537-441     & $2\times 10^3$   & $45.9$ (18.5)  & $42.5$ & 11    & $3.5^{\circ}$  & $ \geq 0.89$   \\ \hline
$^{(*)}$ $^{12}$ PKS 0537-286     & $2\times 10^3$   & $46.5$ (22)  & $43.1$ & 15    & $3^{\circ}$  & $ \geq 0.85$   \\ \hline
$^{12}$ PKS 1508-055  & $2\times 10^3$   & $44.0$ (0.074)  & $43.2$ & 13    & $3^{\circ}$  & $\geq 0.83$   \\ \hline
$^{12}$ PKS  0215+015   & $2\times 10^3$   & $46.1$ (22.5)  & $43.0$ & 13    & $3^{\circ}$  & $\geq 0.80$   \\ \hline
$^{13}$ PKS 2023-07    & $3\times 10^3$ &  $43.9$ (0.080)  &   $42.8$  &  11.8  &  $3^{\circ}$ & $\geq 0.85$  \\ \hline
$^{(*)}$ $^{13}$ 4C 71.07 0836+710     & $3\times 10^3$ &  $44.0$ (0.038)  &   $43.8$  &  14  &  $3^{\circ}$ & $\geq 0.75$  \\ \hline
$^{13}$ S4  1030+61     & $3\times 10^3$ &  $42.9$ (0.038)  &   $42.7$  &  12.2  &  $3^{\circ}$ & $\geq 0.73$  \\ \hline
$^{13}$ PKS 1127-145    & $3\times 10^3$ &  $45.5$ (5)  &   $43.5$  &  11.8  &  $3^{\circ}$ & $\geq 0.66$  \\ \hline
$^{14}$ 4C 38.41 1633+382  & $5\times 10^3$ & $46.2$ (10.6) &   $43.3$ &   12.9    &  $3^{\circ}$  & $\geq 0.73$  \\ \hline
$^{(*)}$ $^{14}$ PKS   0332-403             & $5\times 10^3$ & $45.3$ (5) &   $43.2$ &   10    &  $3^{\circ}$  & $\geq 0.66$  \\ \hline
$^{14}$ PKS 0347-211    & $5\times 10^3$ & $45.8$ (8.6) &   $43.3$ &   12.9    &  $3^{\circ}$  & $\geq 0.64$  \\ \hline
$^{14}$  PKS  2149-306   & $5\times 10^3$ & $46.0$ (8.6) &   $43.6$ &   15    &  $3^{\circ}$  & $\geq 0.64$  \\ \hline
$^{15}$ PKS 2126-158    & $1\times 10^4$  & $46.1$ (8.6)   &  $44.3$  &  14.1    & $3^{\circ}$ & $\geq 0.47$   \\ \hline
$^{(*)}$ $^{15}$ J075303+423130    & $1.3 \times 10^4$  & $44.5$ (0.354)   &  $44.2$  &  15    & $3^{\circ}$ & $\geq 0.40$   \\ \hline
$^{15}$[HB89] 0329-385 & $1.3\times 10^4$ & $44.5$ (4.85)&  $44.2$  &  15  &  $3^{\circ}$    &  $\geq 0.19$   \\ \hline
\end{tabular}
\caption{Mass, radio luminosity, bolometric luminosity measurements together with derived [OIII] luminosity and spin parameter, for various SMBHs. $L_{radio}$ and $L_{\rm OIII}$ are in units of $ \left[ \mathrm{erg \, s^{-1}} \right] $ and in $\log_{10}$ basis. Spin values are predicted by SMFP relation \eqref{eqsmfp}.
The OIII luminosities, Lorentz factor and viewing angles for all AGNs are from \cite{ghis2}, except 15/2 from \cite{Zuo:2014fea}, 15/3 from  \cite{Shemmer:2004ph} and 11/1, 12/2, 13/2, 14/4, 15/1 from \cite{ghis1}. The OIII lines are obtained the relation between OIII lines and  bolometric luminosity given in \cite{Pennell:2017gov}. Radio luminosities are taken from NED.
}
\label{tabestimated}
\end{table}

\subsection{Error Estimation}
While the Spectral Energy Distribution (SED) of the radiation emitted by a BH contains a large wealth of information, it is important to carefully estimate the uncertainties associated with the quantities measured or inferred. They can generally be divided into four categories.

\begin{itemize}

\item {\bf Intrinsic scatter :} This is the residual scatter after the inclusion of the BH spin contribution to the FP variables. As mentioned above, the Spin Modified FP relation decreased the typical scatter from 1 dex to ~0.5 dex, but this scatter cannot be resolved unless we have more information about the physical parameters of the AGN system. There are various contributors to this error, e.g., the thickness of the accretion disk, the viscosity parameter, and the different emission regions.

\item {\bf $L_{\rm bol}-L_{\rm OIII}$ conversion and  measurements :} Typical standard deviation in this conversion is 0.35 dex and we estimate the measurement error to be about 0.5 dex.

\item {\bf Bulk Lorentz factor and viewing angle :} Typical Lorentz factors for blazars lie in the range $[10-20]$, while the viewing angle is generally smaller than a few degrees. The general Doppler boost factor is

\begin{equation}
    \delta_j=\frac{1}{\Gamma_j(1-\beta \times \cos\theta_j)} \, .
    \label{eqforboost}
\end{equation}
As long as $\frac{1}{\theta_j} \ga \Gamma_j$ is satisfied, then the boosting enhancement is set by the Lorentz factor term. Since the luminosity scales as $L_{\rm observed} = \delta_j^2 L_{\rm source}$ for continuous jets, we have $L_{\rm observed} \simeq 4\Gamma^2 \times L_{\rm source}$. The difference between Lorentz boosts given in the above range can lead to an error of 0.3 dex at most.

\item {\bf Luminosity measurements :} Typical uncertainty is 0.3 dex. 

\end{itemize}

\subsection{Lower Bounds for the AGN Spins}

 We employ the error estimations discussed above for the AGNs given in the Table \ref{tabestimated}. The table has multiple AGNs in each mass chunk such that they form a small ensemble. We employ the ensemble error  which is estimated as
\begin{equation}
    \sigma_{ensemble}\equiv \sqrt{\Sigma_i \sigma_i^2} \simeq 0.82
    \label{eqerror}
\end{equation}
Using the relation above together with \eqref{eqsmfp}, we obtain lower bounds on the spin values shown in Table \ref{tabestimated}. The results can be interpreted in the following way : There exists at least one AGN in the given mass chunk spinning faster than the indicated value. For this, we choose the second highest spinning object in the given mass chunk which assures our interpretation at $2-\sigma$ confidence.

We note that the AGNs in the table \ref{tabmeasured} are BHs whose spins are determined by conventional methods, and hence indicated as "measured" category, and the lower bounds on the spins of the AGNs in Table \ref{tabestimated} are predicted via \eqref{eqsmfp} and they are described as "calculated".

\section{Brief Summary of Superradiance}
\label{sec:reviewsuperradiance} 

Superradiance is a phenomenon which non-perturbatively amplifies the near-horizon fields whose wave functions have an angular velocity smaller than the spinning BH \cite{Penrose:1969pc,Misner:1972kx,ZelDovich1972,Starobinsky:1973aij,Teukolsky:1974yv,Detweiler:1980uk}. The ULB fields, which have a Compton wavelength comparable to the size of the event horizon, couple to the BH, and extracts energy and angular momentum from it. Therefore the realization of the superradiance instability requires 

\beq
\omega_b < m \, \Omega_H \, ,
\label{spincond}
\eeq
where $m$ indicates the azimuthal angular number and $\Omega_H (w_H) $ is horizon angular velocity (dimensionless angular frequency), defined as
\beq 
\Omega_H \equiv \frac{a}{2 r_g \left(1+\sqrt{1-a^2}\right)}  = \frac{1}{2r_g} \; w_H \, ,
\eeq
where $r_g=G \times M$ is the gravitational radius. The field is enhanced resonantly and the spin and reducible mass of the Kerr BH decrease until the condition above is not satisfied.

In order for the bosons to deplete the BH spin, on top of the condition \eqref{spincond}, the energy extraction mechanism should act faster than a fundamental scale for BH accretion. Specifically, the characteristic time scale for BH accretion is longer than the instability time scale \cite{SRcond,SRcond1,SRreview}

\beq
 \tau_{BH} > \tau_{SR} 
 \label{timecond}
\eeq

BH growth can be described by the relation
$ \dot M = [(1-{\cal E} (a))/{\cal E} (a)] \; M \; f_{edd}/t_{edd}$, and ${\cal E} (a)$ is the radiative efficiency given in \eqref{eqspinmoddisk}, M is the BH mass with $M \simeq M_{initial} \, \times {\rm e}^{t/\tau_{BH}}$ where a typical  BH growth time scale is given as
\beq
    \tau_{BH} \simeq \frac{{\cal E}}{1-{\cal E}}  \frac{5 \times 10^8}{f_{edd}} \,  \; \mathrm{years}.
\eeq
The accretion rate is indicated by $\dot{M}$, while $\eta$ is the radiative efficiency, $f_{edd}$ indicates the Eddington ratio defined as $L_{bol}/ L_{edd}$, and $t_{edd}= M_{BH}/L_{edd}$. The radiative efficiency factor depends on the spin of the BH, as the Innermost Stable Circular Orbit (ISCO) comes closer for higher prograde orbits.

The instability time scale is indicated as $\tau_{SR}$ and  expressed as \footnote{More details and spin evolution simulations for different accretion rates in \cite{Brito:2014wla}.} \cite{SRcond}

\beq
\tau_{SR} = \frac{log (N_m)}{\Gamma_b} \, ,
\eeq
where $N_m$ is the occupation number for the corresponding state and is expressed as

\begin{equation}
    N_m \equiv \frac{G  M^2 \Delta a}{m} \, ,
    \label{eqoccupationnumber}
\end{equation} 
where $\Delta a$ shows the depleted spin of the BH. $\Gamma_b$ is the field growth rate which shows distinct mass dependencies for different spin fields. For example, spin-2 (tensor) fields have the strongest instability and their particle number grows rapidly, then vectors(spin-1)  and scalars (spin-0) follow in order. Therefore, ruling out scalar degrees of freedom also rules out the same mass range for higher spins (and even a larger range with smaller masses). Here below we report the explicit expressions for the field growth for scalar, vector and tensor fields: \footnote{We note the analytic relations are good enough up to ${\cal O}(1)$ errors, which does not affect our conclusions.} 
\beqa
&& \Gamma_{s=0} \simeq  \frac{1}{12}  \, w_H \left( r_g \, \mu \right)^8 \mu  \nonumber\\
&& \Gamma_{s=1} \simeq 8  \, w_H \, \left( r_g \, \mu \right)^6 \mu \,  
\nonumber\\
&&  \Gamma_{s=2} \simeq    \, w_H \, \left( r_g \, \mu \right)^2 \,\mu   
\label{fieldrate}
\eeqa 

For spin-2 particles, there is also a faster rate  which is proportional to $(r_g \, \mu)^3$ and shown to be valid at small spin values. We take the more conservative rate that is valid for high spins, for details see Ref \cite{Brito:2020lup}.

  \begin{figure}
\centering{ 
\includegraphics[width=0.75\textwidth]{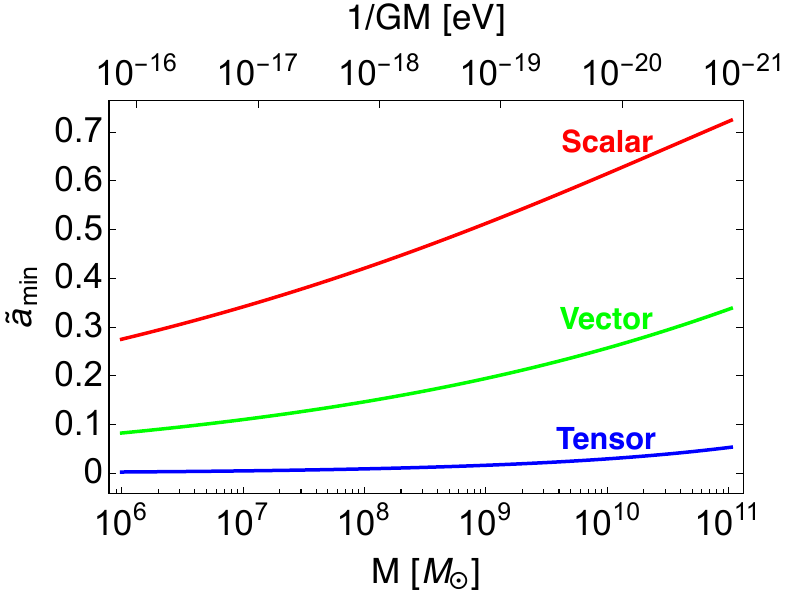}}
\caption{The minimum values of spin as a function of mass in order to constrain the mass parameter of the ultra-light scalar, vector and tensor degrees of freedom, and the typical particle mass in eV on the top of the figure.
}
\label{figminspinvsmass}
\end{figure}

If the conditions \eqref{spincond} and \eqref{timecond} are met, and superradiance is not observed, this constrains the existence of the ultra-light degrees of freedom with a corresponding mass. For a given pair of BH mass and spin there is a minimum boson mass that can be ruled out. A higher spin leads to a larger boson mass range to rule out, until the spin reaches maximum value. This follows from condition \eqref{timecond} on the superradiance. Therefore, we can formulate this case as
\beq
\Omega \geq \mu \geq \Omega_{min}= \mu_{min}\bigg|_{\tau_{BH}=\tau_{SR}} \, 
\label{eqsrrange}
\eeq
If there exists a non-vanishing range for $\mu$ satisfying the above condition and if superradiance is not observed, then one rules out the relevant region. For this it is self-evident that we need $\Omega > \mu_{min}$. While for $a \to 0$ this condition cannot be satisfied, for sufficiently larger values of $a$ this condition can be verified. As $a$ gets larger and reaches its maximum, the mass range satisfying the condition grows. Hence, more rapidly rotating BHs probe a larger mass range.

In Figure \ref{figminspinvsmass}, we obtain the minimum spin value for a given BH mass using \eqref{eqsrrange} by  setting $\tau_{BH}=10^8$ years. Scalar instabilities are smaller, spin-2 types are larger and vectors highest such that even tiny values of spin parameter is enough to set superradiance stage (see also discussions in \cite{SRcond,SRcond1,SRreview,SRconstraints,Stott:2020gjj,SRnumanal}). 

We note that for the rest of the work, we consider $\tau_{BH}$ as a dynamical quantity and as spin grows, radiation efficiency increases and the accretion decreases which influences the dynamical time constant. Moreover, the accretion rate also strongly modifies the time constant, $f_{edd}$ term, which varies considerably for AGNs. For flat spectrum quasars, typically accretion rate is around ${\cal O}(0.1-1)$ Eddington rate. We assume $f_{edd}\simeq 0.5 $ in order to keep our analysis conservative, hence our bounds can be considered as the lowest and robust limits in the given mass ranges.

\section{Implications for Scalar, Vector and Tensor Degrees of Freedoms}
\label{sec:constraints}

\begin{figure}
\centering{ 
\includegraphics[width=0.9\textwidth]{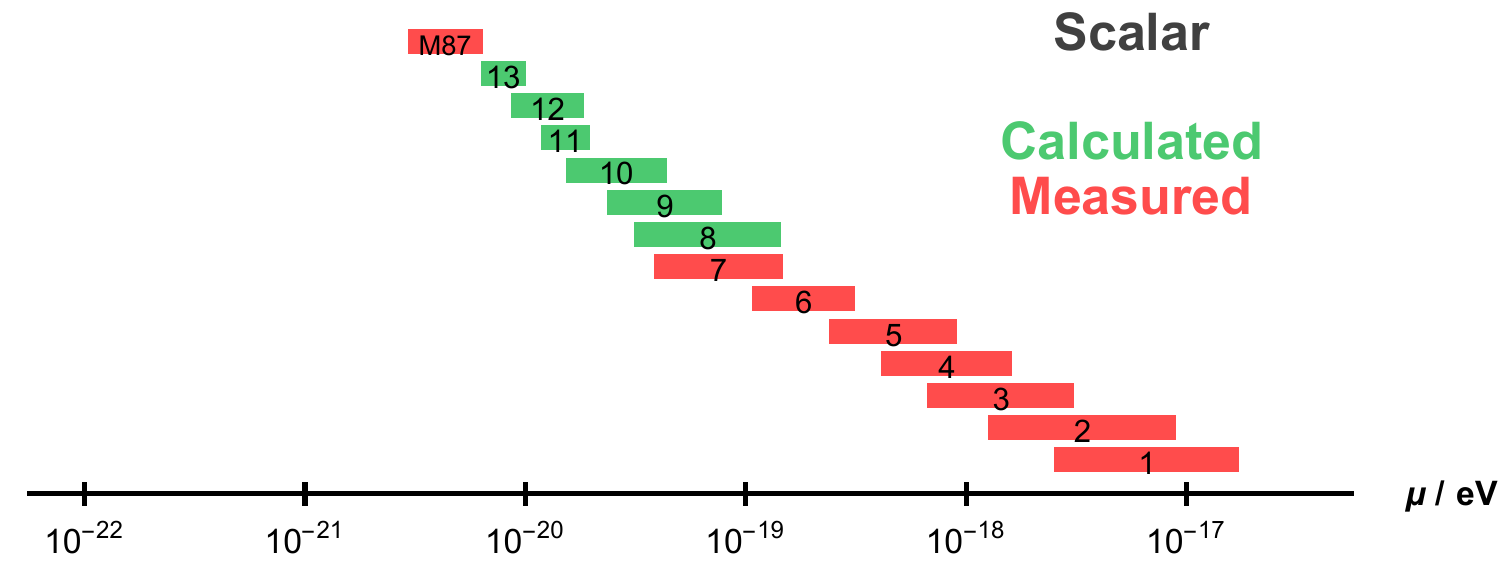}}
\caption{The mass range $\mu$ probed for scalars. Red data are measured (Table 
\ref{tabmeasured}) while green data are calculated (Table \ref{tabestimated}).
}
\label{figscalarpara}
\end{figure}

\begin{figure}
\centering{ 
\includegraphics[width=0.9\textwidth]{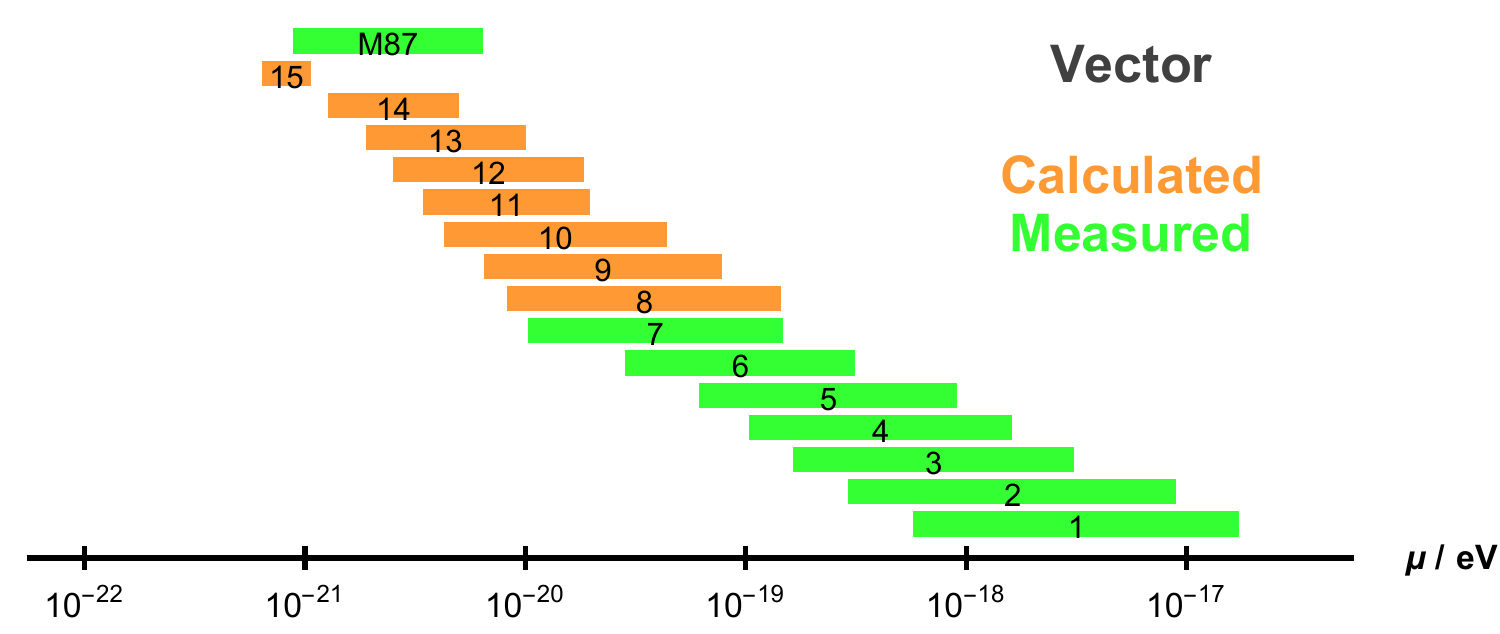}}
\caption{The mass range $\mu$ probed for vectors. Green data are measured (Table 
\ref{tabmeasured}) while the orange data are calculated (Table \ref{tabestimated}).
}
\label{figvectorpara}
\end{figure}

\begin{figure}
\centering{ 
\includegraphics[width=0.9\textwidth]{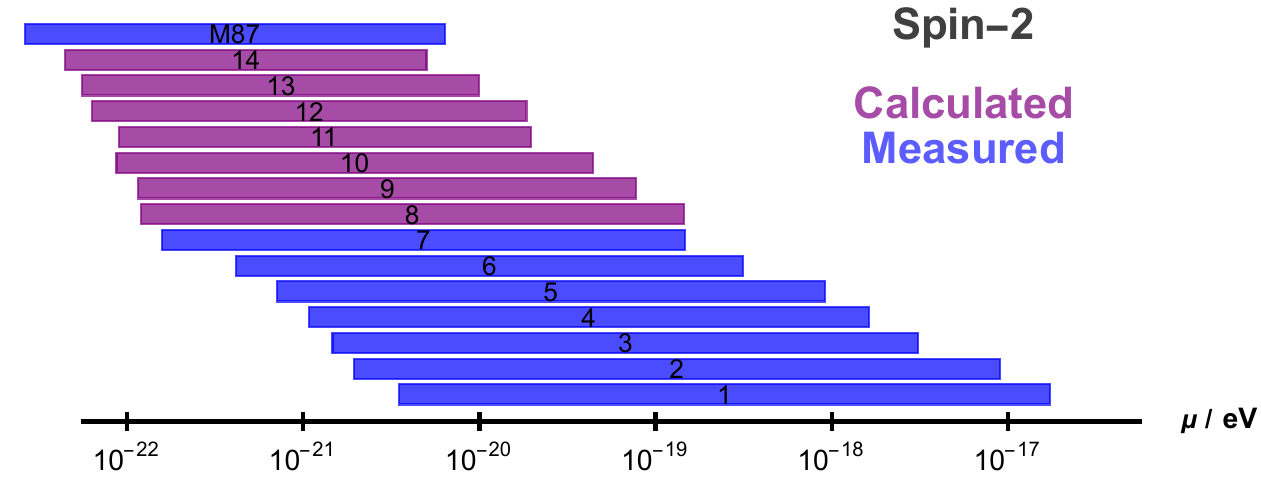}}
\caption{The mass range $\mu$ for tensors. Blue data are measured (Table 
\ref{tabmeasured}), while purple data are calculated (Table \ref{tabestimated}).
}
\label{figtensorpara}
\end{figure}

In the presence of no other interaction, we rule out scalar type ULB \footnote{With the detection of intermediate mass BHs, higher mass ULBs can be constrained \cite{Stott:2020gjj,Wen:2021yhz}.} in the mass range $ 2.9 \times 10^{-21} \, \mathrm{eV}- 1.7 \times 10^{-17} \, \mathrm{eV}$ as shown in Fig. \ref{figscalarpara}. In this section and the following, we note that "measured" AGNs belong to Table \ref{tabmeasured} and "calculated" AGNs belong to Table \ref{tabestimated}. 

In order to constrain scalar boson masses less than $10^{-21}$  eV, larger BH masses are needed. As the BH mass grows, the minimum spin required to probe the parameter space grows as shown in Fig. \ref{figminspinvsmass} (i.e., typically $a \gsim 0.7$), and this requires more precise spin measurements and estimates for BHs heavier than $10^{10} M_\odot$.

For spin-1 and spin-2 degrees of freedom, the interaction between boson and BH is stronger, which translates into a stronger growth rate of the field and a faster superradiance instability. Therefore, much smaller spin parameters are enough to constrain vector and spin-2 degrees of freedom. Assuming other interactions are negligible, for vectors, the probed/ruled out parameter space is $ 6.4 \times 10^{-22} \, \mathrm{eV}- 1.7 \times 10^{-17} \, \mathrm{eV}$ as shown in Figure \ref{figvectorpara}. For spin-2 particles this range is $ 1.8 \times 10^{-21} \, \mathrm{eV}- 1.7 \times 10^{-17} \, \mathrm{eV}$ as shown in Figure \ref{figtensorpara}.  
We have two notes at this point: First, spin-2 constraints under some conditions can be translated to bi-gravity and modified gravity theories as discussed in \cite{spin2SR}. The graviton mass implication, we derive here, is weaker than the bounds derived using the Solar System and galaxy clusters \cite{gravitonbound}. Second, the spin for a BH formed by merger of similar mass slowly spinning BHs is about 0.7; hence any spin larger than this value implies that the rotation is accretion related. Two large mass chunks (14 and 15) given in Table \ref{tabestimated} has at least 1 BH whose spin is larger than 0.66 and 0.4 respectively. However since they are less than 0.7, the bounds derived using them are less robust with respect to smaller mass chunks.

\subsection{Bounds on Axion Decay Constant/Self-Interaction from Superradiance}

An axion field has a typical potential of the form, neglecting higher harmonics, $V=\Lambda^4 (1- \cos \frac{\phi}{f_a} )$, where $\phi$ is the pseudoscalar axion field and $f_a$ denotes the axion decay constant, while the mass is $\mu\simeq \Lambda^2/f_a$. On top of the mass term, by expanding the potential to higher order terms, we obtain self-interaction given by $\frac{\Lambda^4}{f_a^4} \phi^4$. Hence, if the self-interaction term is strong enough, the light degrees of freedom form a condensate and collapse, thus preventing their exponential production via superradiance. Moreover nonlinear and external interactions can also prevent superradiance as discussed in \cite{nosuperradiancebyinteractions1,nosuperradiancebyinteractions2}.

Following Refs. \cite{Yoshino:2012kn,SRcond,SRcond1,SRreview}  for the existence of superradiance in the presence of self-interaction \footnote{For vector and tensor type particles, see for example cubic action for spin-2 \cite{Babichev:2016bxi,spin2SR}. In this work, we focus on scalar self-interaction term. See also \cite{Poddar:2019zoe} for decay constant bounds from compact star binaries.}, we write the condition as

\begin{equation}
\Gamma_{SR} \; \tau_{BH} \, \left(N_{BOSE}/N_{m} \right) > log N_{BOSE} \, ,
\label{eqnovasuperradiance}
\end{equation}
where $N_{m}$ is the occupation number (see \cite{SRcond,SRcond1}) and

\begin{equation}
    N_{BOSE} \simeq 5 \times 10^{94} \frac{n^4}{(r_g \, \mu)^3} \left ( \frac{M}{10^9 \,M_\odot} \right)^2 \left(\frac{f_a}{M_p}   \right)^2
\end{equation}
where the {\cal O(1)} prefactor is obtained with numerical analysis as $\sim5$ in Ref. \cite{Yoshino:2012kn}. Both equations above can be read as follows: As the decay constant $f_a$ decreases, self-interactions become important and $N_{BOSE}$ becomes smaller. This makes satisfying eqn \eqref{eqnovasuperradiance} harder as rate of superradiance have to be larger than ratio $\frac{Nmax}{N_{BOSE}}$.

\begin{figure}
\centering{ 
\includegraphics[width=0.8\textwidth]{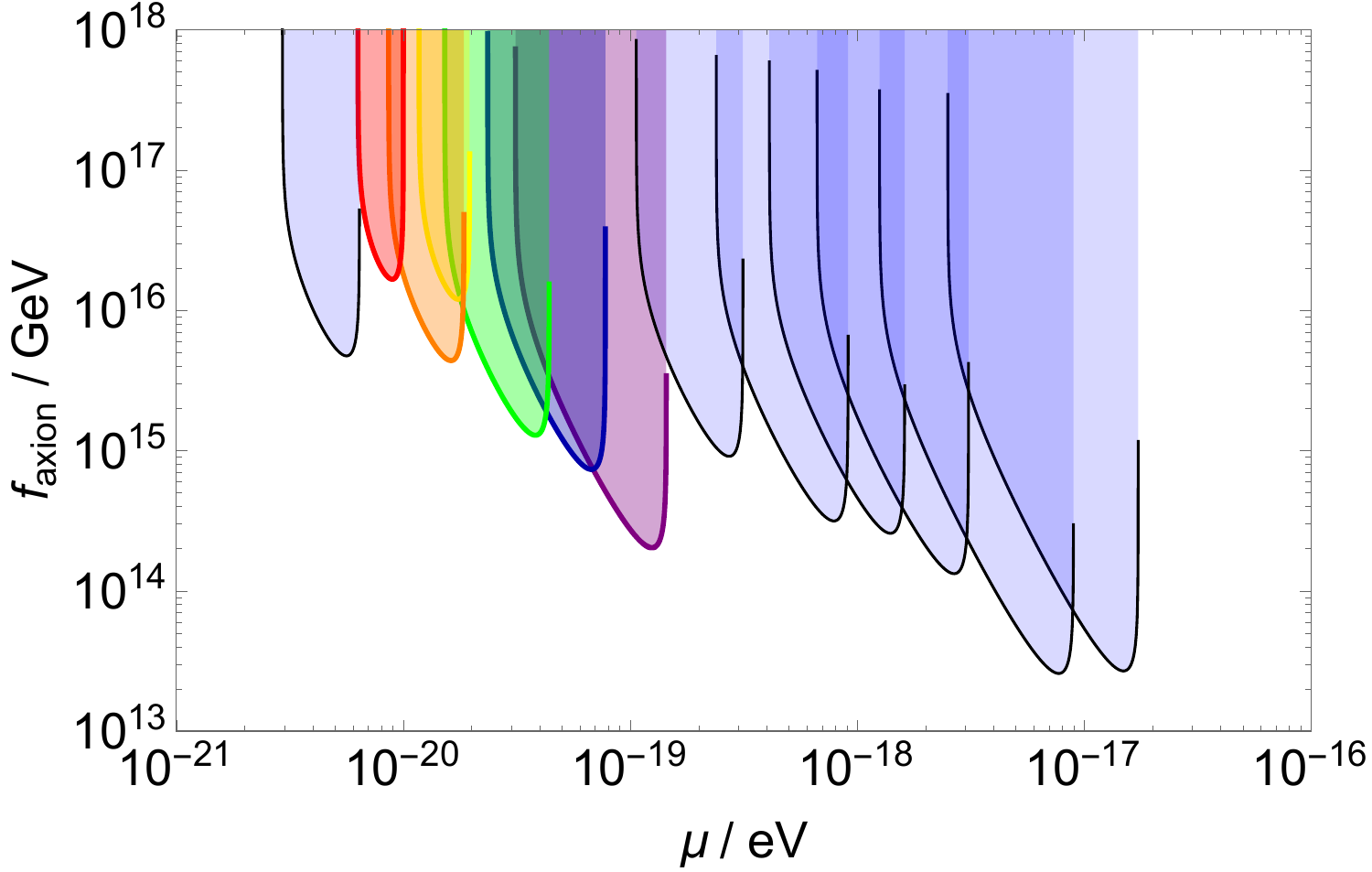}}
\caption{The  bounds on axion decay constant (self-interaction) in the $\mu- f_{axion}$ plane. Rainbow colors correspond to calculated/predicted AGN spins given in Table \ref{tabestimated} and light purple color regions are result of measured AGNs with conventional methods given in Table \ref{tabmeasured}.
}
\label{figselfinteraction}
\end{figure}

\begin{figure}
\centering{ 
\includegraphics[width=0.8\textwidth]{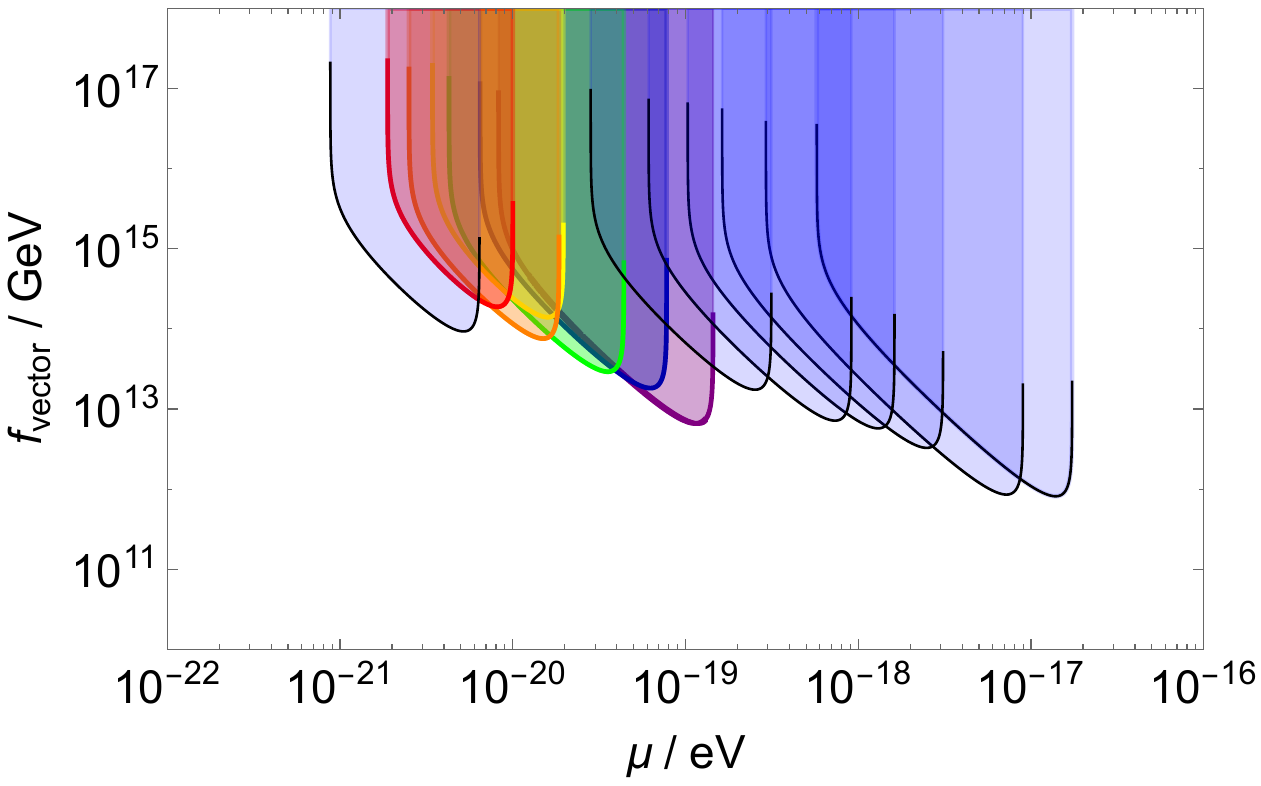}}
\caption{The  bounds on vector decay constant (self-interaction) in the $\mu- f_{vector}$ plane assuming vector field is Abelian which results in similar to scalar case. Rainbow colors correspond to calculated/predicted AGN spins given in Table \ref{tabestimated} and light purple color regions are result of measured AGNs with conventional methods given in Table \ref{tabmeasured}.
}
\label{figvectorselfinteraction}
\end{figure}

\begin{figure}
\centering{ 
\includegraphics[width=0.8\textwidth]{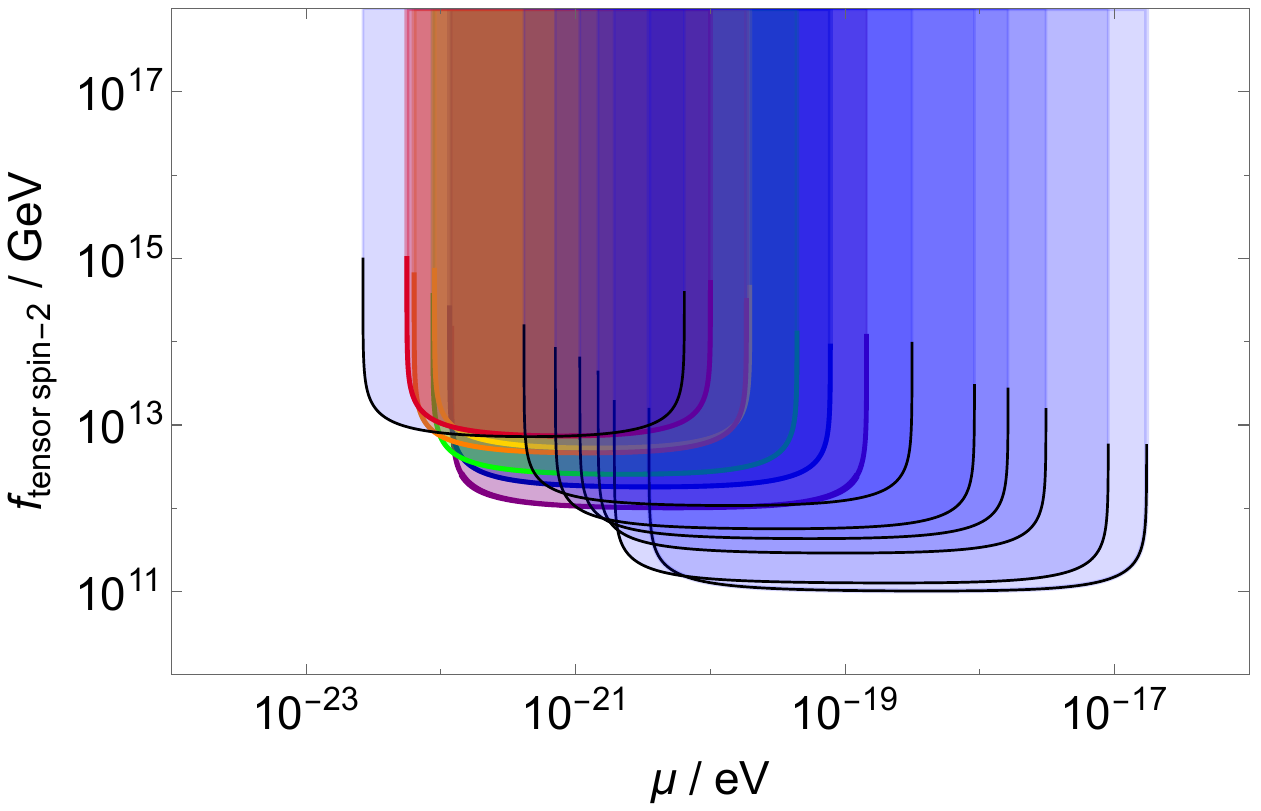}}
\caption{The  bounds on spin-2 decay constant (self-interaction) in the $\mu- f_{vector}$ plane assuming tensor field has similar self-interaction form with the scalar case. Rainbow colors correspond to calculated/predicted AGN spins given in Table \ref{tabestimated} and light purple color regions are result of measured AGNs with conventional methods given in Table \ref{tabmeasured}.
}
\label{figtensorselfinteraction}
\end{figure}

These points are summarized in Figure \ref{figselfinteraction}. If the superradiance is not observed, then either there is no such particle with the corresponding mass ranges, or non-observation of the superradiance is prohibited by self(or external) interactions. The self-interactions preventing ULB to enter in superradiance requires smaller values of decay constant, hence non-existence of superradiance can introduce upper bounds on the decay constants for different particle masses. Since we have spinning heavy BHs, then the allowed decay-constant is indicated by the white regions below colored regions. Blue color is for the BHs whose spin is measured from the spectrum, and rainbow color region is produced by BHs whose spins are predicted via 
SMFP given in \eqref{eqsmfp} with error estimate and shown in Table \ref{tabestimated}.

According to recent studies \cite{Gruzinov:2016hcq,Baryakhtar:2020gao} the nonlinear interactions between different energy levels are more important in most of the parameter space instead of bosenova formation. The physical scale for decay constant that saturates spin-down is found to be similar order. Although the decay constant values are inferred similar order, the physical processes are different. The resulting torque on the black hole is found as \cite{Gruzinov:2016hcq,Baryakhtar:2020gao}
\begin{equation}
{\cal T} \sim a^{3/2} \; \alpha^7 \; \mu \;  \frac{f^2}{\mu^2} \equiv \frac{dL}{dt} \sim \frac{G M^2 a}{T_{spindown}}
\end{equation}
Hence timescale for spindown is roughly given as
\begin{equation}
T_{spindown} \sim \frac{G M^2 }{a^{1/2} \alpha^7 \; \mu  \;  \frac{f^2}{\mu^2}}
\end{equation}
It is remarkable to observe that other than log dependent piece, bosenova bounds and nonlinear self interaction bounds  on $f$ are within O(1) level different and can be given as
\begin{equation}
\frac{f}{M_p}  \sim  10^{-4}  \left(\frac{10^{10} \mathrm{yr}}{\tau_{\mathrm{BH}}}\right)^{\frac{1}{2}}\left(\frac{10^{-13} \mathrm{\, eV}}{\mu}\right)^{\frac{1}{2}} \\
\left(\frac{0.01}{\alpha}\right)^{\frac{5}{2}}\left(\frac{0.9}{a}\right)^{\frac{1}{2}}
\end{equation} 
The effect of the nonlinear interactions and bounds on self-interactions can be incorporated for a more complete picture, which we leave for future studies.


\subsection{ Are Ultra Light Bosons Dark Matter?}
\label{secdarkmatter}

\begin{figure}
\centering{ 
\includegraphics[width=0.8\textwidth]{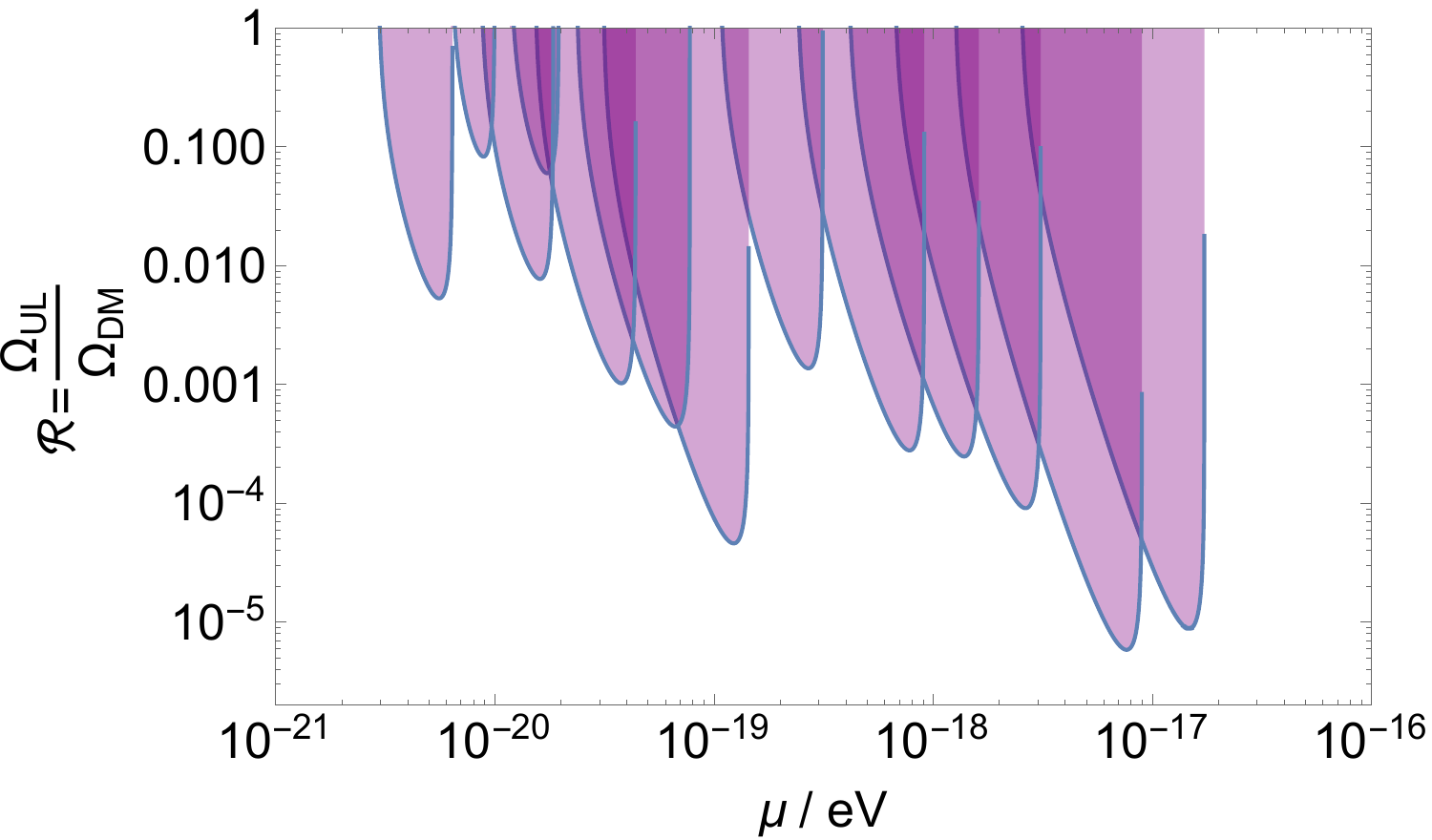}}
\caption{The constraints on the fraction of dark matter composed of ultra light scalar particles.}
\label{figdarkmatter}
\end{figure}

\begin{figure}
\centering{ 
\includegraphics[width=0.8\textwidth]{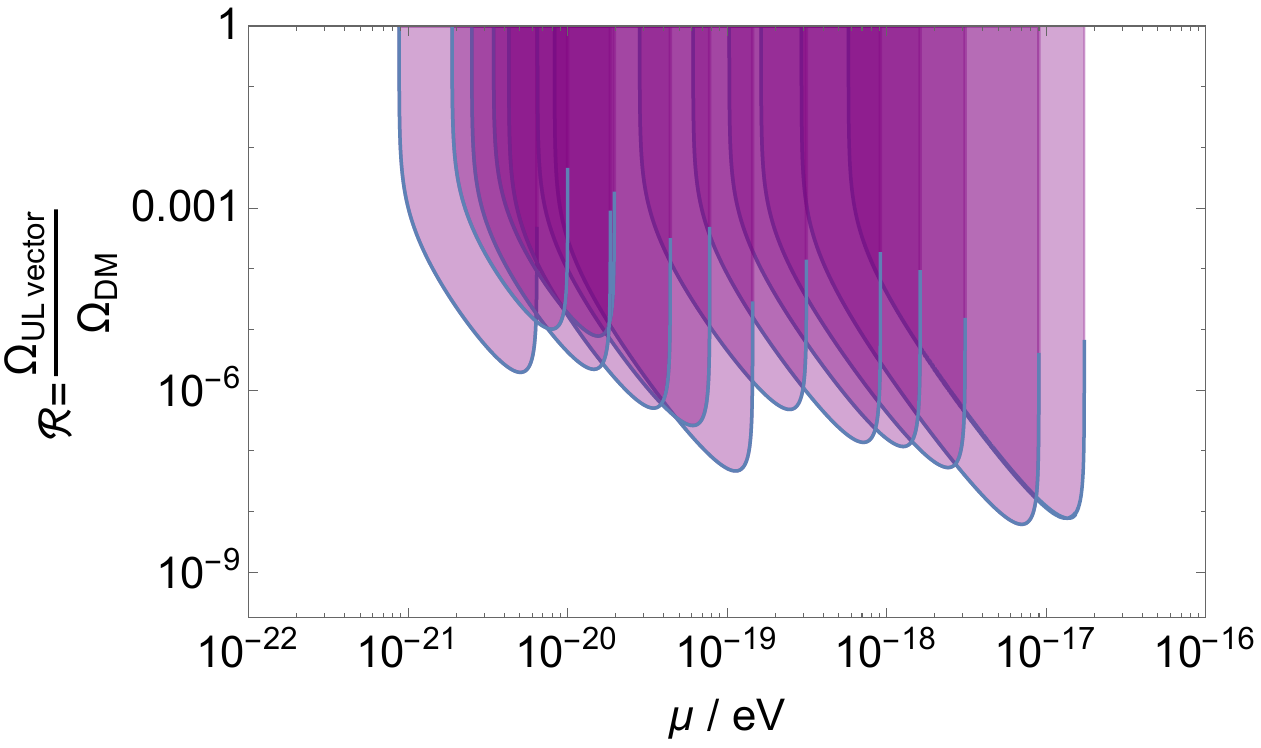}}
\caption{The constraints on the fraction of dark matter composed of ultra light vector particles.}
\label{figdarkmattervector}
\end{figure}

\begin{figure}
\centering{ 
\includegraphics[width=0.8\textwidth]{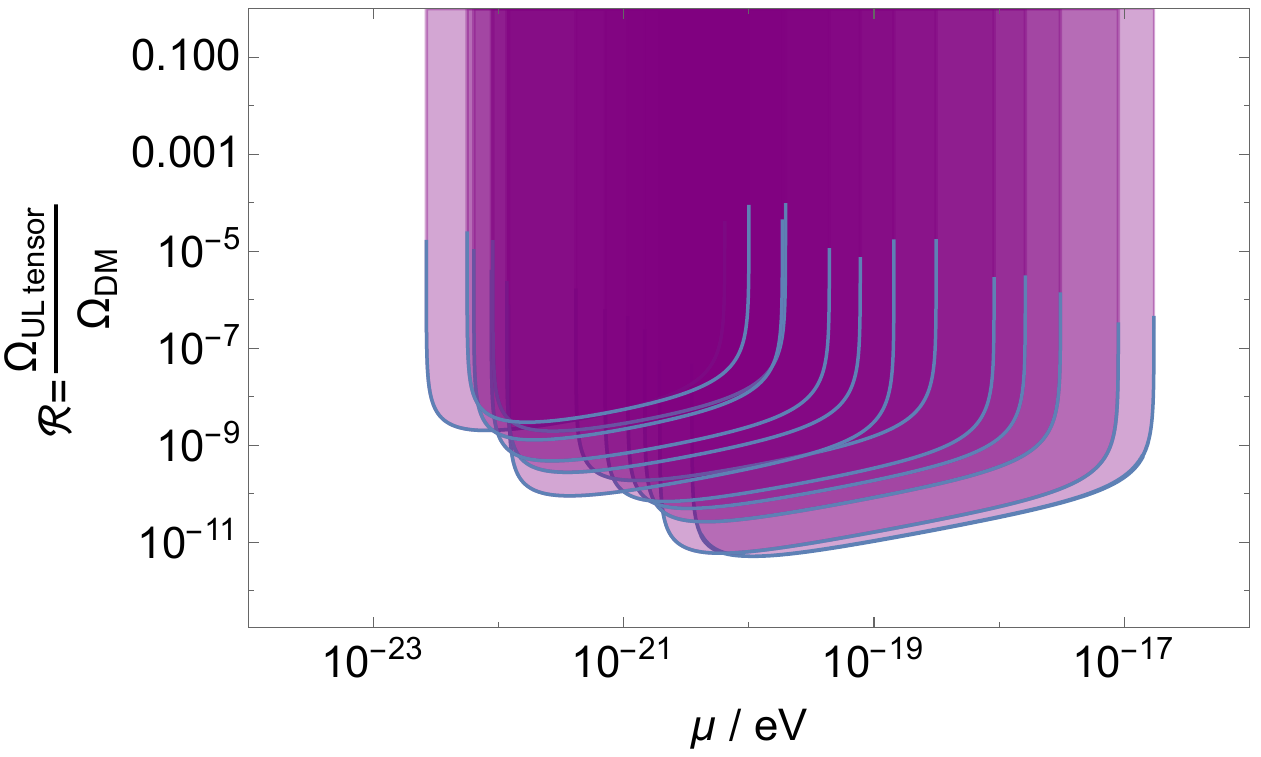}}
\caption{The constraints on the fraction of dark matter composed of ultra light tensor particles.}
\label{figdarkmattertensor}
\end{figure}

The scalar, vector and spin-2 ULB could make up the dark matter \cite{Hu:2000ke,Hui:2016ltb,Marsh:2015xka,Marzola:2017lbt}. We focus on the contribution of scalar, vector and tensor ultra light particles to the energy density.

Assume we have a light degree of freedom.  It does not evolve as the Hubble parameter is larger than its mass, namely until $H\sim m$. 

\begin{equation}
    \rho=3H_{move}^2 M_p^2 \sim T_{move}^4 \Rightarrow H_{move}\sim m\sim T_{move}^2/M_p
\end{equation}

In order for this matter to be dark matter, it needs to have same energy density with radiation at redshift about 3,000 or $T_{eq}\sim eV$. Since light particles decay like dust/non-relativistic species after $m>H$, the energy density at matter-radiation equality should be equal to radiation around eV. Assuming $\phi/f_a\sim O(1)$, this gives 

\begin{eqnarray}
   \frac{\rho_{axion}(t_{eq})}{\rho_{radiation}} &\sim&  \frac{\Lambda^4}{T_{move}^4} \frac{ R_{eq}}{R_{move}} \sim 1 \Rightarrow m^2 f_a^2\sim T_{move}^3 T_{eq} \nonumber\\ \nonumber\\
   && \Rightarrow m^2 f_a^2 \sim ( m \cdot M_p)^{3/2} T_{eq}
\end{eqnarray}
here R is scale factor. They all result in \cite{Hui:2016ltb}
\begin{equation}
       {\cal R}_{Ultra \, Light}  \equiv \, \frac{\Omega_{UL}}{\Omega_{DM}} \sim \, \left( \frac{\mu}{10^{-21} \, \mathrm{eV}}\right)^{1/2} \left( \frac{f_a}{10^{17} \, \mathrm{GeV}} \right)^2 \,.
\end{equation}
Using the constraints obtained for self-interaction in the previous subsection, we investigate the fraction of dark matter that can be contributed by ULB particles in Figure \ref{figdarkmatter}. We find that in the range, $ {\rm a \, few} \times  10^{-21}- 10^{-17} \, \mathrm{eV}$, scalar light particles can be at most $10\%$ of all dark matter,  vector light particles can be at most $10^{-6}$ of dark matter energy budget and spin-2 particles  can be at most $10^{-9}$ of all dark matter.

\section{Summary and Conclusions}
\label{sec:conc}

In this work we employed the SED of AGNs to infer the BH spin. We improve the previously established SMFP relation by exchanging X-ray luminosity with the [OIII] line, and incorporating the relativistic corrections such as Lorentz factor and viewing angle of the jet. Thanks to the remarkable results obtained, we were able to put lower bounds on the spin of AGNs and, consequently, probe the properties of hypothetical Ultra-Light Bosons (ULB) which interact with AGN if their Compton wavelength is comparable with the horizon size. In result of superradiance instability, part of angular momentum and energy of BH can be transfered to ULB. We examined the unexplored parts of the parameter space of the ULB with 9 AGNs whose spins are measured via conventional methods and given in Table \ref{tabmeasured}, and 29 quasars whose minimum spin is inferred via \eqref{eqsmfp} and \eqref{eqerror}  and results given in Table \ref{tabestimated}. We obtain the following constrains, assuming negligible self/external interactions
\begin{itemize}
\item $2.9 \times 10^{-21} \, {\rm eV} \, - \,  1.7 \times 10^{-17} \, {\rm eV}$ for scalars (spin-0 particles),
\item $6.4 \times 10^{-22} \, {\rm eV} \, - \,  1.7 \times 10^{-17} \, {\rm eV}$ for vectors (spin-1 particles),
\item $2.6 \times 10^{-23} \, {\rm eV} \, - \,  1.7 \times 10^{-17} \, {\rm eV}$ for tensors (spin-2 particles).
\end{itemize}
It is important to note that superradiance instability of BHs to ultra light scalar, vector and spin-2  perturbations exists in the negligible external/self interaction conditions. Therefore, employing the fact that the superradiance is not observed at given mass scales, we  obtain the upper bounds on the axion decay constant (inversely proportional to self-interaction) that prevents ultra light particles from entering the superradiance instability. Finally we derive the maximum fraction of scalar, vector (Abelian) and tensor ULB contributing to dark matter, by combining its mass and the corresponding decay constant. We find that ULB can constitute at most I) For scalars : 10\%  of the dark matter in $10^{-21}\, \mathrm{eV} < \mu < 10^{-19} \, \mathrm{eV}$ range, and 0.01-1\% of dark matter in $ 10^{-19}\, \mathrm{eV} > \mu>10^{-17}\, \mathrm{eV}$ range, II) For vectors : $10^{-6}$  of the dark matter in $10^{-21}\, \mathrm{eV} < \mu < 10^{-19} \, \mathrm{eV}$ range, and  $10^{-8} - 10^{-6} $ of dark matter in $ 10^{-19}\, \mathrm{eV} > \mu>10^{-17}\, \mathrm{eV}$ range, III) For tensors : $10^{-9}$  of the dark matter in $10^{-22}\, \mathrm{eV} < \mu < 10^{-17} \, \mathrm{eV}$ range.

\vspace{3mm}
%
 \acknowledgments

  We thank Asimina Arvanitaki, Kfir Blum, Richard Brito, Vitor Cardoso, \"Unal Ertan, Xiaohui Fan, Ely Kovetz, David Marsh, Paolo Pani and Surjeet Rajendran  for the discussions on AGNs, accretion discs, ultra-light particles and superradiance, and for the feedback on the draft, and particularly  Payaswini Saikia for the explanations on optical FP and Federico Urban for the ultra light particles and their interactions.  C.\"U. dedicates this work to gentle souls of \.Inan Av\c{s}ar, Furkan Celep, Saadet Harmanc{\i} and Sibel \"Unli, and all the young people who lost their lives due to mobbing, bullying and harassment.  C.\"U. thanks his family for their support, Tichá Kavárna and Middle East Technical University Physics Department
 for their hospitality, members of CEICO team in the Institute of Physics of Czech Academy of Sciences
  and Ben Gurion University personnel
 for their help and support during the transition period. C.\"U. is supported by European Structural and Investment Funds and the Czech Ministry of Education, Youth and Sports (Project CoGraDS - CZ.$02.1.01/0.0/0.0/15\_003/0000437$).
 F.P. and A.L. acknowledge support from the Black Hole Initiative at Harvard University, which is funded by grants from the John Templeton Foundation and the Gordon and Betty Moore Foundation. F.P. acknowledge support from Clay Fellowship administered by the Smithsonian Astrophysical Observatory.




\end{document}